\documentclass[prb,twocolumn,aps,showpacs,fixfloats]{revtex4}
\usepackage{graphicx}
\usepackage{bm}
\usepackage{amsmath, amsthm, amssymb} 
\usepackage{color}
\usepackage{times}
\usepackage{amsfonts}
\newcommand{\s}{\scriptscriptstyle}
\newcommand{\bq}{\begin{equation}}
\newcommand{\ee}{\end{equation}}
\newcommand{\eps}{\varepsilon}

\begin{document}

\title{Plasmon spectrum and plasmon-mediated energy transfer in a multi-connected geometry}

\author{L. Shan, E. G. Mishchenko, and M. E. Raikh}

\affiliation{Department of Physics and Astronomy, University of Utah, Salt Lake
City, UT 84112, USA}

\begin{abstract}
Surface plasmon spectrum of a metallic hyperbola can be found analytically
with the separation of variables in elliptic coordinates. The spectrum
consists of two branches: symmetric, low-frequency  branch, $\omega <\omega_{\s 0}/\sqrt{2}$,
and antisymmetric high-frequency branch,  $\omega >\omega_{\s 0}/\sqrt{2}$, where $\omega_{\s 0}$ is the bulk plasmon frequency. The frequency width  of the plasmon band increases with decreasing the angle between the asymptotes of the hyperbola.
For the simplest multi-connected geometry of two hyperbolas separated by
an air spacer the plasmon spectrum contains two low-frequency branches
and two high-frequency branches. Most remarkably, the lower of two low-frequency
branches exists at $\omega \rightarrow 0$, i.e., unlike a single hyperbola,
it is ``thresholdless."  We study how the complex structure of the plasmon
spectrum affects the energy transfer between two emitters located on the surface
of the same hyperbola and on the surfaces of different hyperbolas.

\end{abstract}

\maketitle

\section{Introduction}

The two central
issues of the contemporary plasmonics are manipulation and focusing of light on subwavelength scales, and plasmon-mediated energy transfer, see recent reviews
Refs.~\onlinecite{SchullerReview,Bozhevolny1,Bozhevolny2,Bozhevolny3}.

With regard to the first issue, strong confinement of optical
fields at small scales is accompanied by their orders-of-magnitude enhancement.
This enhancement can be used, e.g., to boost nonlinear effects or to image and
detect\cite{StockmanSensing} small objects.
At the core of the second issue is the light-matter interaction.
The effect of plasmon-supporting interface on a group of emitters located nearby\cite{SPPEnhanced,lifetime1,lifetime2,lifetime3,GovorovEnhanced,Cooperative,Zhenya,Lukin1,Lukin2}
is twofold. Firstly, it strongly modifies the radiative lifetimes of individual emitters. Secondly, virtual plasmon exchange facilitates the dipole-dipole interaction between emitters.

A metallic strip or a wire of finite thickness can serve as a plasmonic
waveguide. The key approach to a plasmon field focusing utilizes tapering, i.e. gradual narrowing of a waveguide towards one end\cite{original1,original2,original3,Stockman,AfterStockman0,AfterStockman1,AfterStockman2,JAP,JPhys,Express,Plasmonics,Blair}.
 Field confinement at the end of a tapered metal wire waveguide was demonstrated experimentally\cite{Bozhevolny3,Focusing0,FocusingExperiment,FocusingExperiment1}.

Current advances in plasmonics are related to engineering of progressively more complex plasmonic  structures\cite{PlasmonicLense,PlasmonicCrystal,PlasmonPhotonicCrystal,structure,strip,Zayats}.
Then it is natural to extend the theoretical study of plasmonic waveguides
to these complex, in particular, multi-connected, geometries, which contain several {\em disconnected} metal-air boundaries. It can be expected that these geometries provide additional control over plasmon fields.
Studies of plasmons  in multi-connected geometries should encompass calculation of the spectra of plasmonic modes as  well as investigation of interaction of dipole emitters mediated by such modes.

\begin{figure}[t]
\centerline{\includegraphics[width=70mm,angle=0,clip]{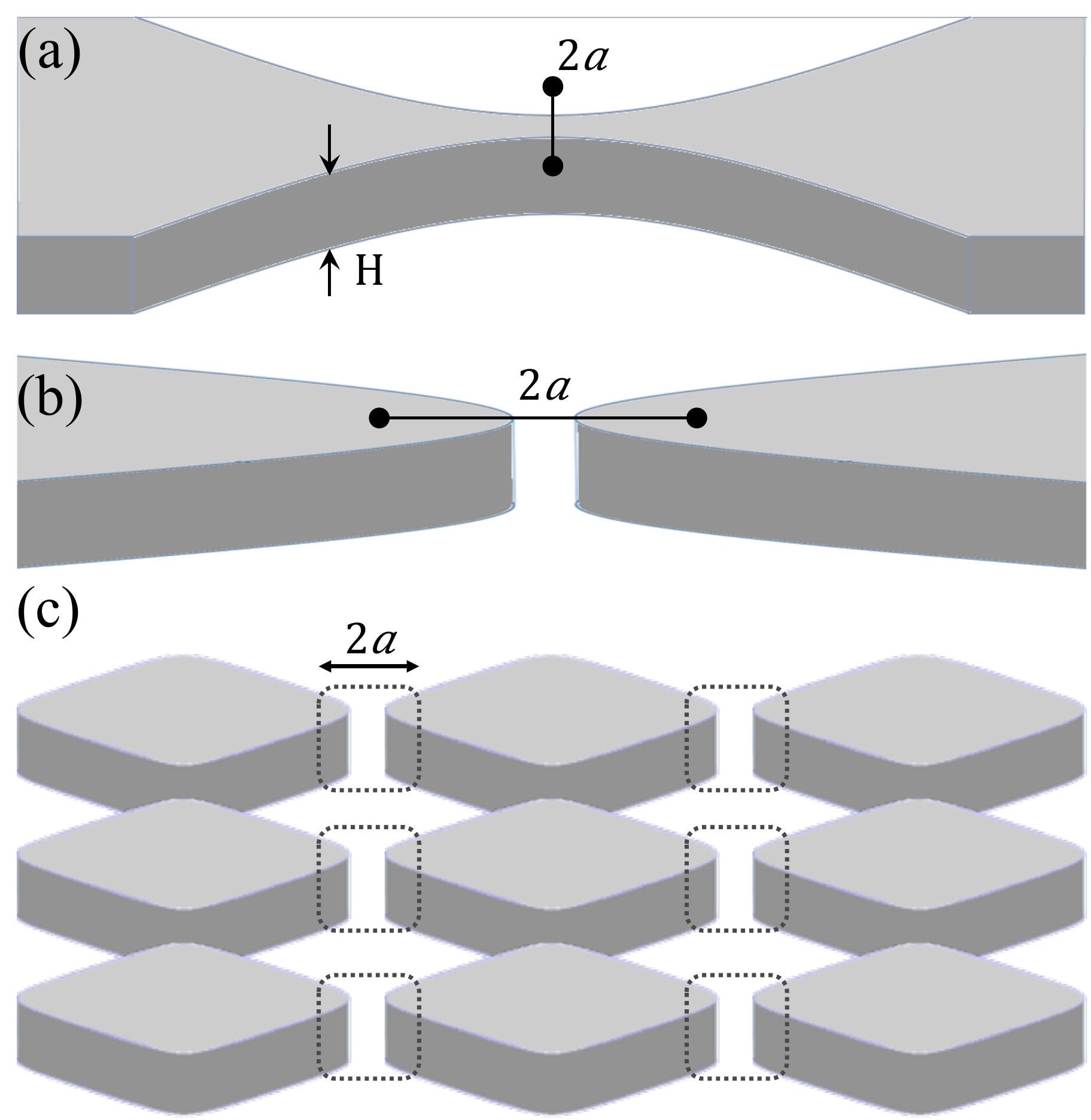}}
\caption{Simplest examples of multi-connected plasmonic structures: a) ``neck" and b) ``gap."
The plasmon spectrum of these structures can be found analytically if metallic surfaces are confocal hyperbolas; $2a$ is the distance between the foci. c) Schematic view of a plasmonic array\cite{array}. The field distribution in the array is determined by the points of contact of metallic islands.  }
\label{F1}
\end{figure}

Below we consider the simplest example of a multi-connected geometry illustrated in Fig.~\ref{F1}.
Geometries of a metallic ``neck," Fig.~\ref{F1}a, and of  two tapers separated by
a narrow air ``groove," Fig.~\ref{F1}b,
are dual to each other. Both geometries possess a characteristic length scale $\sim a$, which is the size of the gap or the width of the constriction. This scale, being much smaller than all other scales in plasmonic structure, see Fig.~\ref{F1}c,  will therefore
determine the plasmon spectrum and the plasmon field distribution of the entire structure.

We take advantage of the fact that the problem of calculation of the plasmon spectrum
in geometry shown in Fig.~\ref{F1} can be solved exactly if the metallic surfaces are hyperbolic.
A natural unit of momentum (wave number) is $q\sim a^{-1}$; the meaning of $q$ in the absence of translational symmetry will be clarified below.
One advantage of the analytical approach over numerical methods applied to concrete sets of parameters \cite{NumericalHyperbolas} is that an analytical solution allows to establish general properties of the
plasmon spectrum.

\begin{figure}[t]
\centerline{\includegraphics[width=85mm,angle=0,clip]{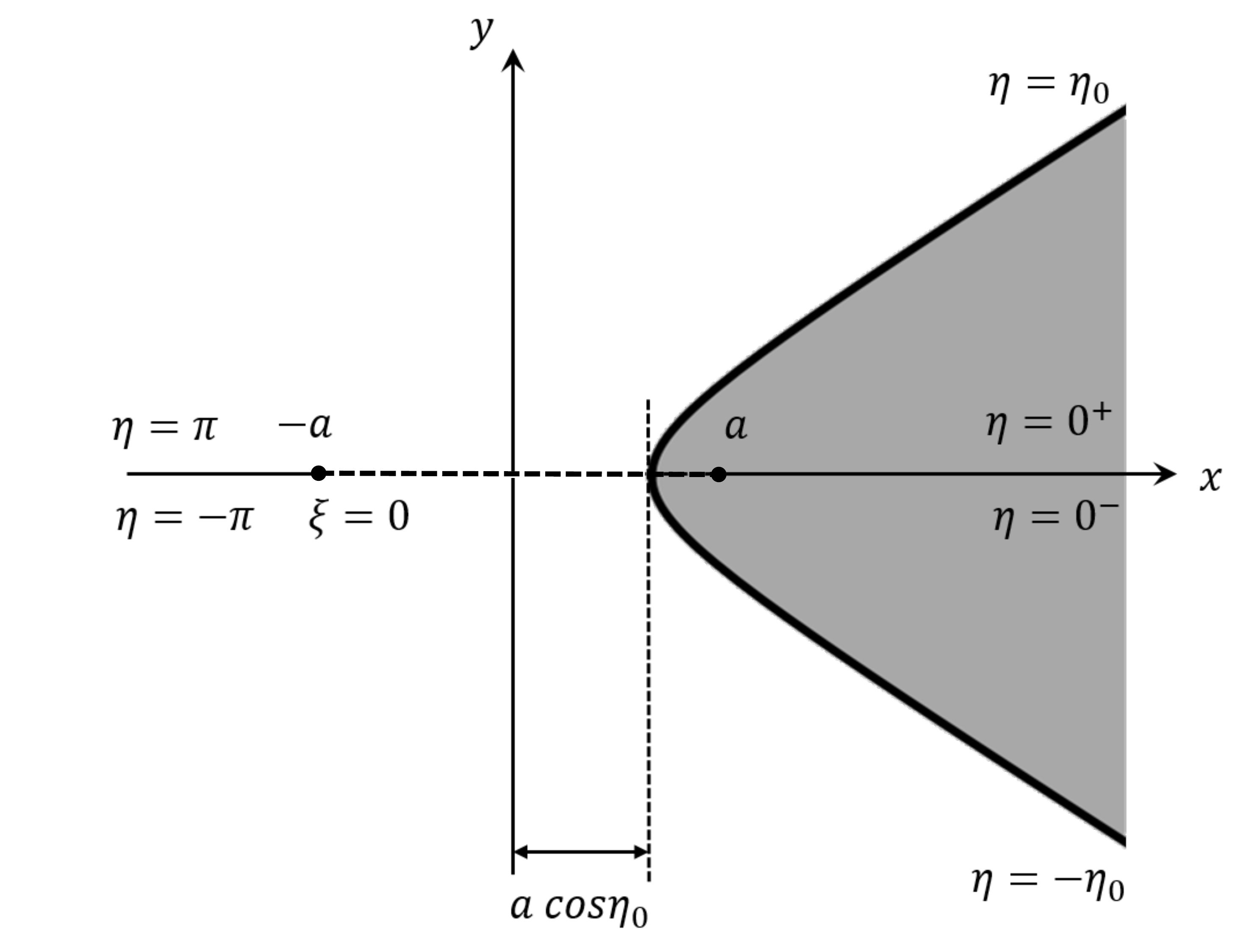}}
\caption{Geometry of a single hyperbola. In elliptic coordinates
$\xi,\eta$, defined by
 Eq. (\ref{hyperbolic coordinates}), the metal with dielectric constant,
 $\varepsilon(\omega)$, occupies the domain
 $\left(-\eta_{\s 0},\eta_{\s 0}\right)$.
 Outside the hyperbola there is air with
$\varepsilon =1$.}
\label{F2}
\end{figure}

We compare the analytical results for two hyperbolas, Fig.~\ref{F1}, to the
geometry of a single hyperbola. For a single hyperbola, the plasmon spectrum,
resembles the spectrum of a finite-width metallic strip. It consists
of two branches:
the low-frequency branch, $\omega <\omega_{\s 0}/\sqrt{2}$,
where $\omega_{\s 0}$ is the bulk plasmon
frequency, corresponds to a symmetric mode, while the high-frequency branch,
$\omega>\omega_{\s 0}/\sqrt{2}$, corresponds to antisymmetric mode.
Adding the second hyperbola, Fig. \ref{F1}, leads to the emergence of two additional branches in the plasmon
spectrum. The frequency of the additional symmetric mode is lower than that for a single hyperbola,
while the frequency of the additional antisymmetric mode is higher a single hyperbola's mode.
Since the plasmons  mediate the energy exchange between the emitters close to the surface,
the complex plasmon structure in a multi-connected geometry leads to a nontrivial frequency
dependence of this exchange, which we study analytically.

The paper is organized as follows.  In Sec. II we present a detailed analysis
of the plasmon spectrum and the field distribution of the plasmon modes in a
 single hyperbola geometry. This analysis is then extended to the geometry of two hyperbolas
is Sec. III.  Excitation of plasmon modes by an emitter located at the metal boundary
in a multi-connected geometry is studied in Sec. IV. Section V concludes the paper.

\section{Plasmon spectrum of a single hyperbola}
\subsection{Elliptic Coordinates}
We start with  a single-hyperbola geometry, Fig. \ref{F2}. The boundary of a metal is
defined by the equation
\begin{equation}
\label{boundary1}
\Bigl(\frac{x}{\cos\eta_{\s 0}}\Bigr)^2-\Bigl(\frac{y}{\sin\eta_{\s 0}}\Bigr)^2=a^2.
\end {equation}
The metal is described with a dielectric permittivity,
\begin{equation}
\label{epsilon}
\eps(\omega)=1-\frac{\omega_{\s 0}^2}{\omega^2},
\end{equation}
and occupies the inner part of the hyperbola, which approaches the asymptotes $y=\pm~x\tan\eta_{\s 0}$ at large $x$. Outside the hyperbola there is air with $\eps =1$.

Surface plasmons correspond to the solution of the Laplace equation, $\nabla \left(\varepsilon\nabla \Phi \right)=0$, for the electrostatic potential, $\Phi(x,y)$, which propagates along the boundary of the hyperbola and decays away from the boundary.  The variables in the 2D Laplace equation can be separated in  elliptic coordinates
\begin{equation}
\label{hyperbolic coordinates}
x=a\cosh\xi\,\cos\eta,~~~~y=a\sinh\xi\,\sin\eta,
\end{equation}
where $\xi$ is positive and $\eta$ changes in the interval $-\pi <~\eta~<~\pi$. The Lam\'e coefficients are the same for both coordinates, $h_\xi=h_\eta =\sqrt{\cosh^2{\xi}-\cos^2{\eta}}$.
In the new coordinates the Laplace  equation,
\begin{equation}
\label{laplace's equation}
\nabla^2\Phi =
\frac{1}{h^2_\xi}\left(\frac{\partial^2\Phi}{\partial \xi^2}+\frac{\partial^2\Phi}{\partial \eta^2}\right) =0,
\end{equation}
has the structure similar to its usual Cartesian form. The solutions of Eq.~(\ref{laplace's equation})
should satisfy the boundary conditions at $\eta=\eta_{\s 0}$ (upper half of the hyperbola) and at $\eta= -\eta_{\s 0}$
(lower half of the hyperbola): the potential $\Phi$ and the normal component of the displacement field, $D_n=\varepsilon h^{-1}_{\xi}{\partial \Phi}/{\partial \eta}$,
should be continuous at this boundary.

\begin{figure}[t]
\centerline{\includegraphics[width=85mm,angle=0,clip]{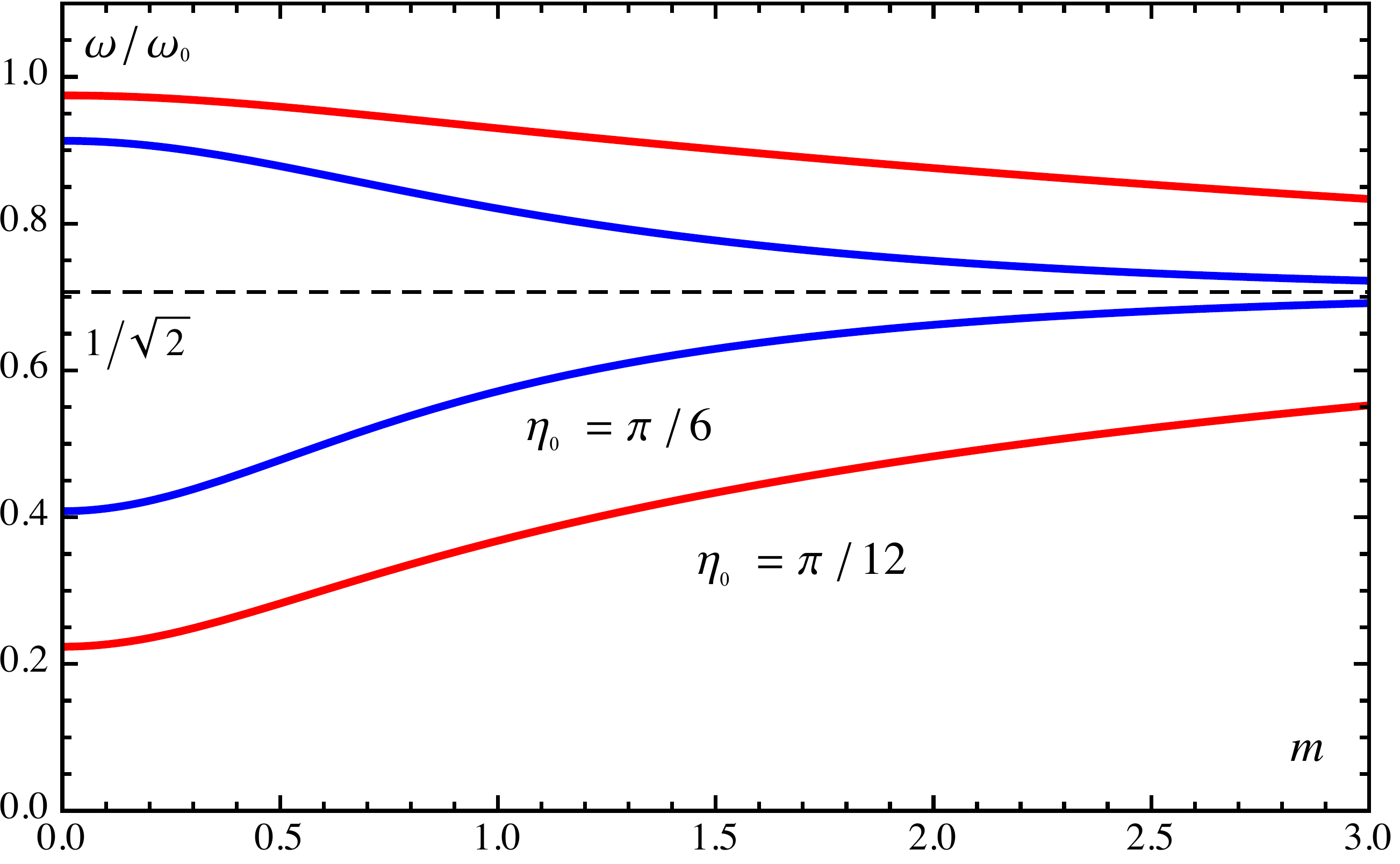}}
\caption{(Color online) Plasmon spectrum of a single hyperbola  for two opening angles
$2\eta_{\s 0}=\pi/10$ (red) and $2\eta_{\s 0}=\pi/3$ (blue) is plotted from Eq.~(\ref{single symmetric}) for a symmetric mode and Eq.~(\ref{single antisymmetric}) for antisymmetric mode. At large wavenumber, $m$, both branches approach the flat surface plasmon frequency, $\omega_{\s 0}/\sqrt{2}$.}
\label{F3}
\end{figure}

\begin{figure}[t]
\centerline{\includegraphics[width=90mm,angle=0,clip]{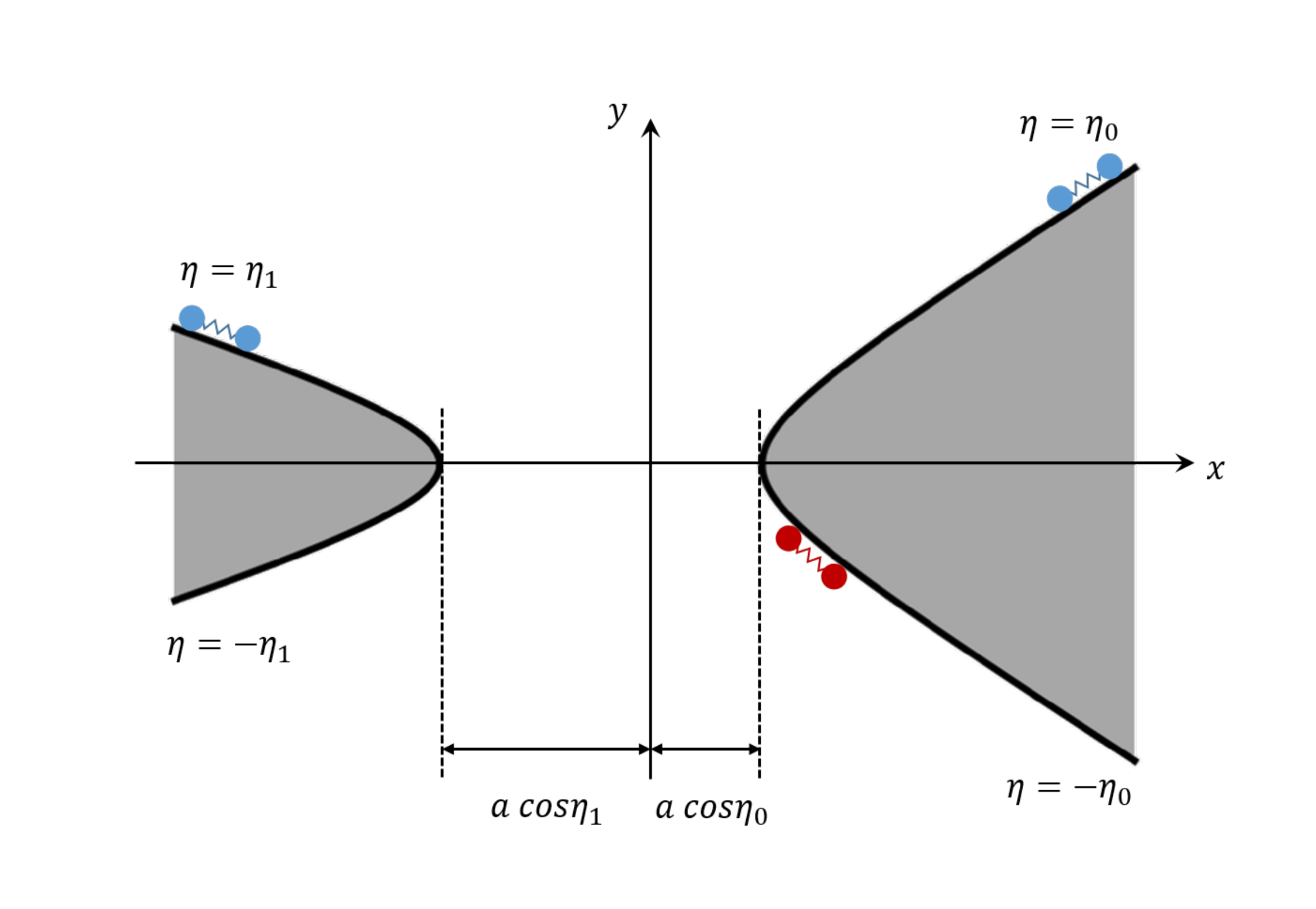}}
\caption{The geometry of two hyperbolas with the opening angles $2\eta_{\s 0}$ and $2\eta_{\s 1}$.
A dipole emitter close to the tip and polarized along the interface excites the dipoles located to the left   and to the right from the tip at distances much bigger than the focal distance.}
\label{asy}
\end{figure}

\begin{figure}[t]
\centerline{\includegraphics[width=85mm,angle=0,clip]{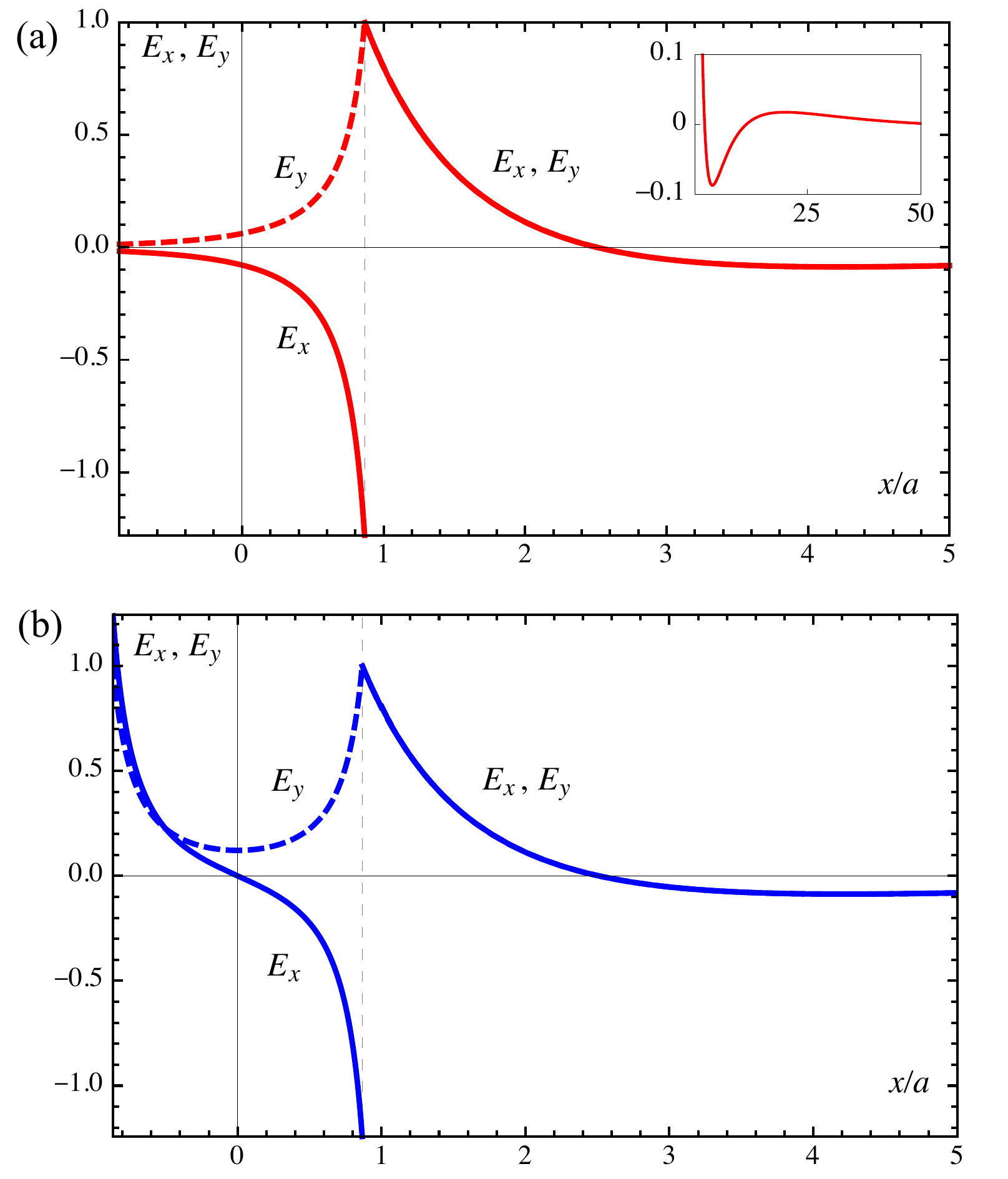}}
\caption{(Color online) (a) Distribution of  $E_x$ (for a symmetric mode) and
$E_y$ (for antisymmetric mode), the components of electric field along the
$x$-axis, is plotted from Eqs. (\ref{field1})-(\ref{field3}) for the wavenumber $m=2$ and the opening angle $2\eta_{\s 0}=\pi/3$. The inset shows the large-$x$ oscillating tail of $E_x$.
(b) The same as (a) for two hyperbolas. }
\label{fields}
\end{figure}

\subsection{Plasmon Dispersion}

The geometry in Fig. \ref{F2} is symmetric with respect to the change of sign of the $y$-axis.
Correspondingly, the solutions of Eq.~(\ref{laplace's equation}) can be
classified into symmetric (even), $\Phi_s(x,y)=\Phi_s(x,-y)$, and antisymmetric (odd),  $\Phi_a(x,y)=-\Phi_a(x,-y)$. In hyperbolic coordinates the reversal of the sign of $y$ amounts to the change $\eta \to -\eta$, so that
$\Phi_s(\xi,\eta)=\Phi_s(\xi,-\eta)$ and $\Phi_a(\xi,\eta)=-\Phi_a(\xi,-\eta)$.

Consider first the symmetric modes. For these modes, a general solution of  Eq. (\ref{laplace's equation})
inside the hyperbola, $|\eta|<\eta_0$, which propagates along $\xi$ and grows with $\eta$ from $\eta=0$
towards the boundary, has the form
\begin{equation}
\label{1}
\Phi_s(\xi,\eta)=A\Bigl[\exp(im\xi)+\alpha\exp(-im\xi)\Bigr]\cosh(m\eta).
\end{equation}
In order to satisfy the continuity of the potential at $\eta =\pm \eta_0$, the corresponding solution outside the hyperbola, $|\eta|>\eta_0$, must have the same $\xi$-dependence.
It should also decay away from the boundary. This specifies the form of $\Phi_s(\xi,\eta)$ in the air:
\begin{equation}
\label{2}
\Phi_s(\xi,\eta)=B\Bigl[\exp(im\xi)+\alpha\exp(-im\xi)\Bigr]\cosh m(\pi-|\eta|).
\end{equation}
By matching
$\Phi_s(\xi,\eta)$ and $\varepsilon \partial \Phi_s/\partial \eta$ at
the boundary $\eta=\pm\eta_{\s 0}$, we obtain the dispersion equation for the symmetric modes
\bq
\label{2'}
\eps(\omega) \tanh{(m\eta_{\s 0})} \coth{[m(\pi -\eta_{\s 0})]}= -1.
\ee

At first glance it appears that the  constant $\alpha$ in Eqs.~(\ref{1})-(\ref{2}) can be arbitrary. However, from the requirement that the function $\Phi_s(\xi,\eta)$ is finite and continuous it follows that $\alpha =1$. Indeed, for any $\alpha \ne 1$ the $\xi$-component of electric field, $E_\xi =- h_\xi^{-1} {\partial \Phi_s}/{\partial \xi}$, diverges when $\xi \to 0$ and $\eta \to 0$, since the Lam\'e coefficient $h_\xi$ vanishes there.
We thus conclude that $\Phi_s(\xi) \propto \cos \left(m\xi\right)$.

Consideration for antisymmetric modes proceeds along the same lines. It is not difficult to see that the solution
$\cos{(m\xi)}\sinh(m\eta)$ must be discarded as discontinuous across the region of the $x$-axis, where $-a <x<a$ and $\xi=0$: the odd character of $\sinh{(m\eta)}$ with respect to  $\eta \to -\eta$ implies that it can only by multiplied by a function of $\xi$ that vanishes at $\xi=0$. This suggests the following form of the potential, $\Phi_a(\xi,\eta)$,
for antisymmetric modes
inside the hyperbola

\begin{equation}
\label{3}
\Phi_a(\xi,\eta)=C\sin (m\xi)\sinh(m\eta).
\end{equation}
Correspondingly, the potential outside the hyperbola, which is odd with respect to $y$ and  matches the $\xi$-dependence of Eq.~(\ref{3}) has the form
\begin{equation}
\label{4}
\Phi_a(\xi,\eta)=D\sin (m\xi)\sinh m(\pi-|\eta|).
\end{equation}
The resulting dispersion equation for antisymmetric modes is
\bq
\label{5}
\eps(\omega) \coth{(m\eta_{\s 0})} \tanh{[m(\pi -\eta_{\s 0})]}= -1.
\ee
From Eqs. (\ref{2'}) and (\ref{5}) we can express the plasmon frequencies
in terms of the dimensionless wavenumber, $m$,
\begin{equation}
\label{single symmetric}
\omega_s(m)=\omega_{\s 0}\Biggl\{ \frac{\sinh(m\eta_{\s 0})\cosh[m(\pi-\eta_{\s 0})]}{\sinh(m\pi)}\Biggr\}^{1/2},
\end{equation}
\begin{equation}
\label{single antisymmetric}
\omega_a(m)=\omega_{\s 0}\Biggl\{ \frac{\cosh(m\eta_{\s 0})\sinh[m(\pi-\eta_{\s 0})]}{\sinh(m\pi)}\Biggr\}^{1/2}.
\end{equation}
Examples of dispersions Eqs.~(\ref{single symmetric}), (\ref{single antisymmetric}) are shown in Fig.~\ref{F3}. Below we list some general properties of these dispersions:

\vspace{1mm}
(i) A hyperbola reduces to a plane for $\eta_{\s 0}=\pi/2$. Then  Eqs.~(\ref{single symmetric}), (\ref{single antisymmetric}) reproduce the expected result, $\omega_s(m)=\omega_a(m)=\omega_{\s 0}/\sqrt{2}$,
for the frequency of a dispersionless  surface plasmon.

\vspace{1mm}

(ii) The spectra Eqs.~(\ref{single symmetric})-(\ref{single antisymmetric})
of the symmetric and antisymmetric modes satisfy the ``sum rule" relation,
\begin{equation}
\omega_s^2(m)+\omega_a^2(m)=\omega_{\s 0}^2.
\end{equation}

\vspace{1mm}

(iii) In the long-wavelength limit, $m\to 0$, the threshold frequencies are
\begin{equation}
\label{longwave}
\omega_s(0) =\omega_{\s 0}\Bigl(\frac{\eta_{\s 0}}{\pi}\Bigr)^{1/2},~~~~ \omega_a(0) =\omega_{\s 0}\Bigl(1-\frac{\eta_{\s 0}}{\pi}\Bigr)^{1/2}.
\end{equation}
The fact that $\omega_s(0)$ goes to zero for small $\eta_{\s 0}$ could be expected, since at small
$\eta_{\s 0}$ the hyperbola is effectively a metallic layer with  zero thickness, for which the dispersion of the longitudinal surface plasmon  does not have a threshold\cite{Stern}.
On the other hand, $\omega_a$ approaching $\omega_{\s 0}$ for $\eta_{\s 0}\rightarrow 0$
can be interpreted by noticing that the oscillations of the electron density, accompanying this plasmon,
are normal to the surface; such oscillations must have  the  bulk-plasmon frequency $\omega_0$.

\vspace{1mm}

(iiii) Another notable property of the dispersions Eqs. (\ref{single symmetric})-(\ref{single antisymmetric})
is their {\it duality}, namely,
\begin{equation}
\label{duality}
\omega_s(\pi-\eta_{\s 0},m)=\omega_a (\eta_{\s 0},m).
\end{equation}
Qualitative interpretation of this relation can be given for small $\eta_{\s 0}$, when Eq.~(\ref{duality}) relates the spectrum in of a sharp metal ``edge'' with the spectrum of a narrow ``funnel'': it suggests that the frequency of a soft symmetric plasmon of the ``edge'' coincides with the frequency of an antisymmetric plasmon in a ``funnel." Indeed, once the plasmon frequency is low, the absolute value of $\varepsilon(\omega)$ is big. This, in turn,  implies that the normal component of the oscillating electric field in the metal is small.
In the funnel geometry, this small field amplitude can be realized only when  {\em opposite} charges
accumulate on the two surfaces of the funnel. The reason is that, for small-angle funnel, such antisymmetric arrangement  is similar to the charges of a parallel-plate capacitor; this distribution of charges ensures that the field of the parallel-plate capacitor does not extend outwards. On the other hand, as we mentioned above, the low-frequency plasmon in a sharp-edge geometry is longitudinal, which
corresponds to {\em symmetric} amplitudes of the charge-density fluctuations for the two surfaces.

\subsection{Field Distribution}

\begin{figure}[t]
\centerline{\includegraphics[width=90mm,angle=0,clip]{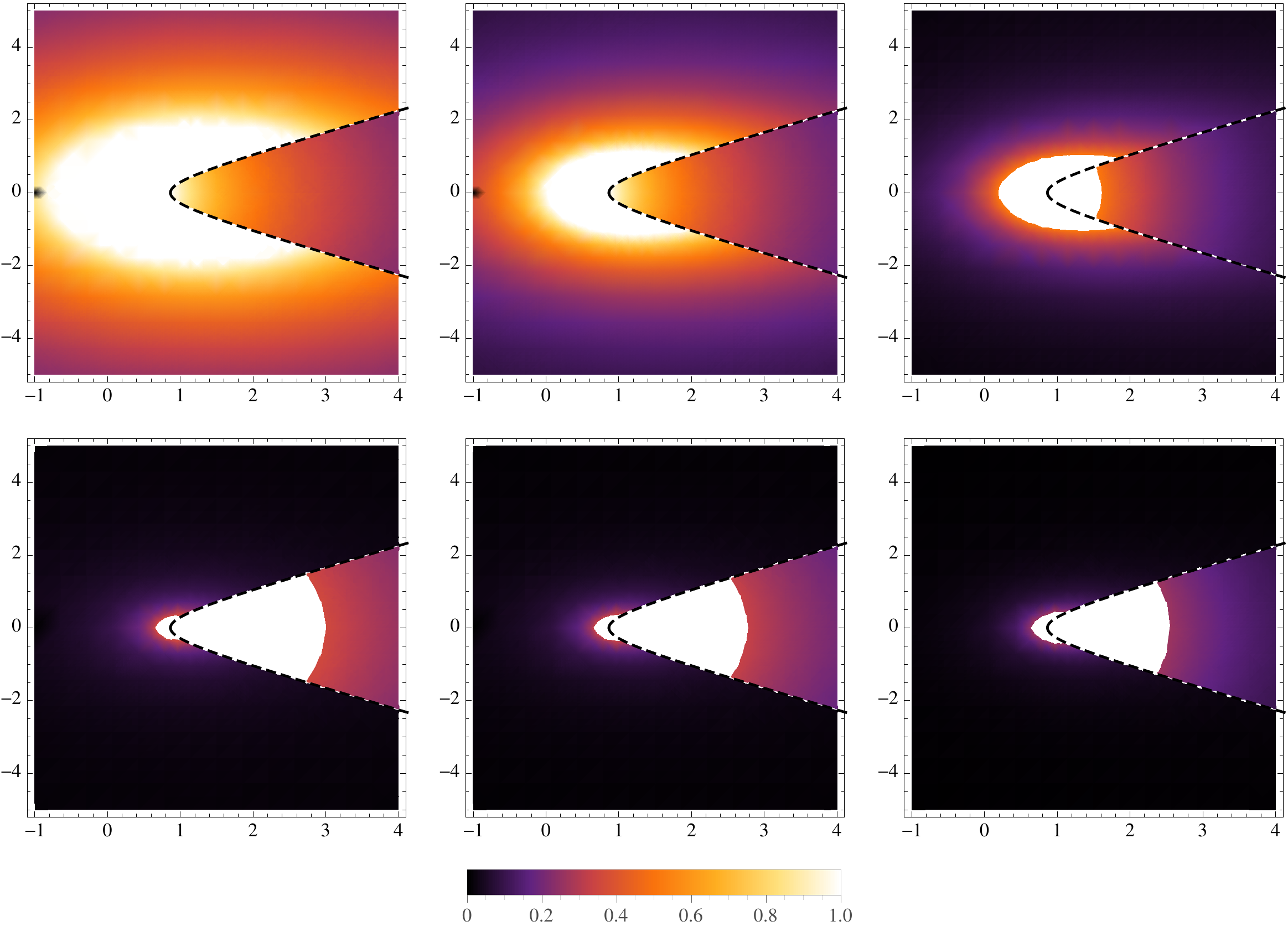}}
\caption{(Color online) Density plot of the field intensity of
the plasmon modes for a single hyperbola with the opening angle $2\eta_{\s 0}=\pi/3$.
The upper row corresponds to a symmetric mode, for which the potential distribution is determined
by Eqs. (\ref{1}) and (\ref{2}). The lower row corresponds to an antisymmetric mode with the potential
described by Eqs.~(\ref{3}) and (\ref{4}). Left, central, and right panels correspond to the wavenumbers, $m=0.1$, $m=0.4$, and $m=0.7$, respectively. With increasing $m$ the  plasmon frequencies approach
$\omega_{\s 0}/\sqrt{2}$, while the field concentrates near the metal-air surface.   }
\label{F4}
\end{figure}

\begin{figure}[t]
\centerline{\includegraphics[width=90mm,angle=0,clip]{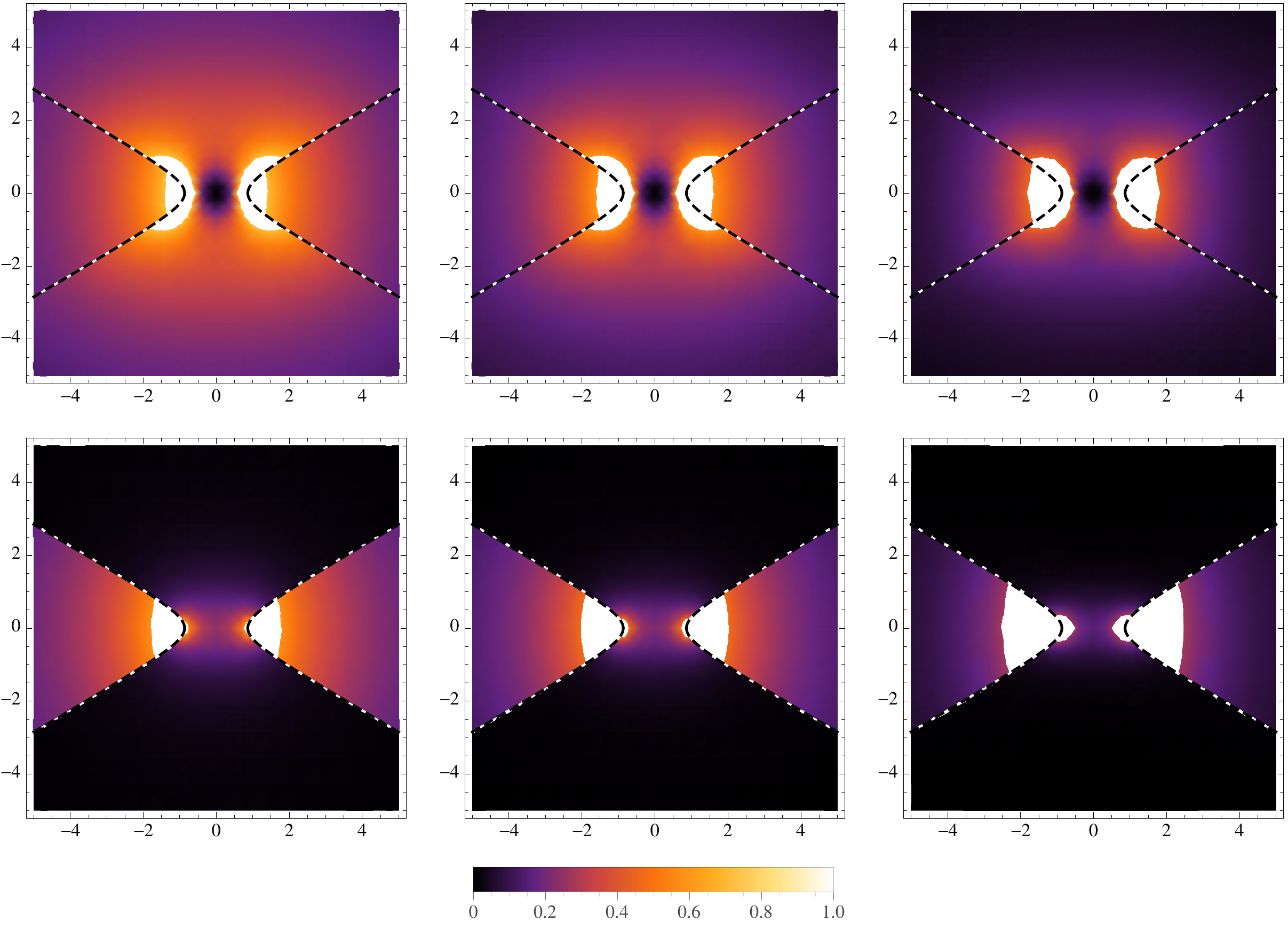}}
\caption{(Color online) Density plot of the field intensity of
the plasmon modes in the geometry of two symmetric hyperbolas with the same opening angle, $2\eta_{\s 0}=\pi/3$, as in Fig. \ref{F4}.
The upper row corresponds to a symmetric mode for which the potential distribution is determined
by Eqs. (\ref{double symmetric}) and (\ref{solution}).
Left, central, and right panels correspond, respectively, to the wavenumbers, $m=0.1$, $m= 0.4$, and $m=0.7$, the same as in Fig. \ref{F4}. While the fields of individual
hyperbolas are disconnected along the $x$-axis, they overlap along the $y$-axis.
The lower row corresponds to antisymmetric modes with the same $m$-values. The field of individual hyperbolas overlap, predominantly, along the $y$-axis.  }
\label{F5}
\end{figure}

\begin{figure}[t]
\centerline{\includegraphics[width=90mm,angle=0,clip]{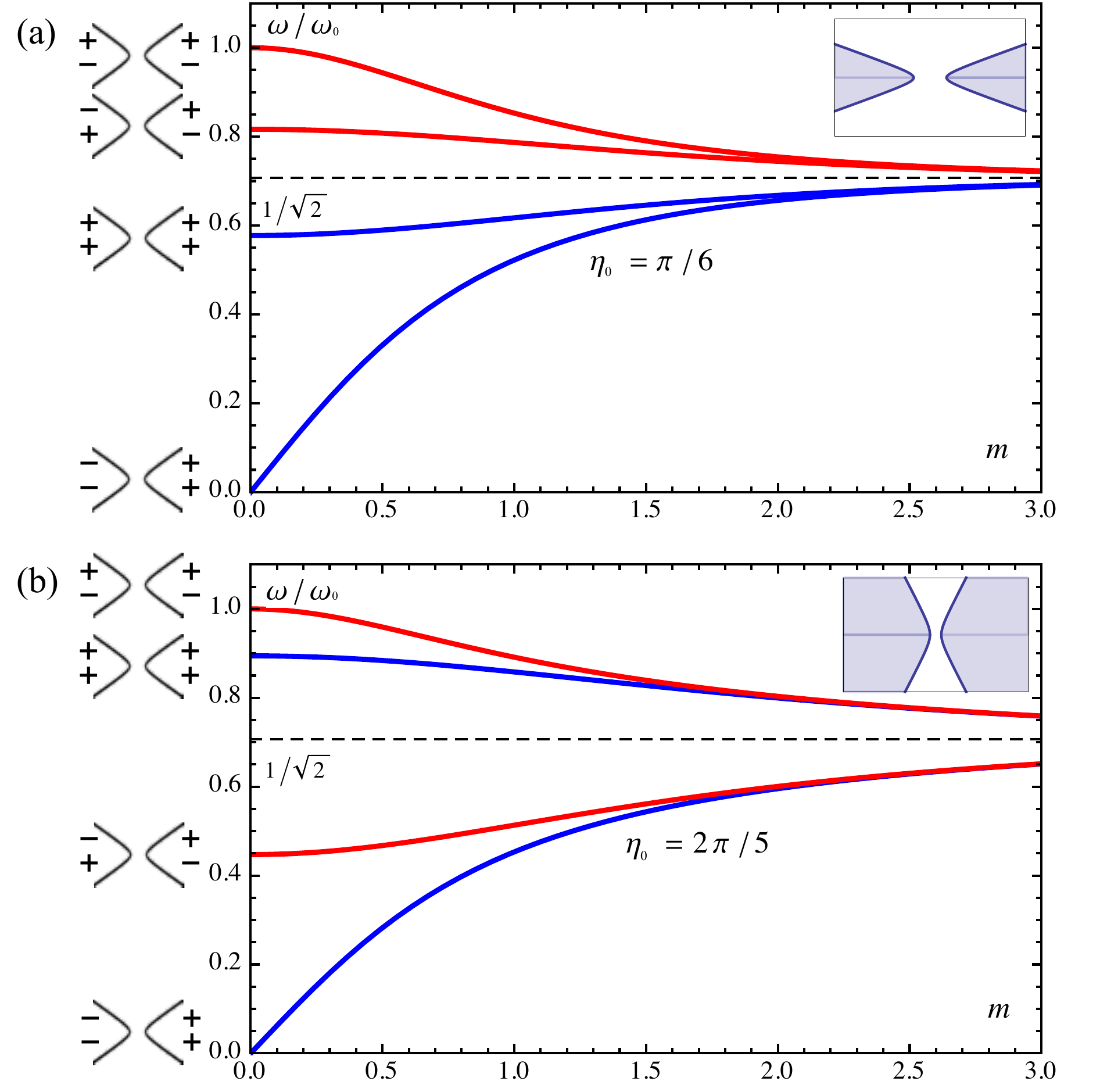}}
\caption{(Color online) Upper panel: Comparison of the plasmon spectra in the geometry of two hyperbolas for two values of the opening angle. For $\eta_{\s 0}=\pi/6<\pi/4$ the frequencies
of both  symmetric modes (blue) are smaller than the flat surface plasmon frequency $\omega_{\s 0}/\sqrt{2}$, while the frequencies
of both antisymmetric modes (red) are bigger than $\omega_{\s 0}/\sqrt{2}$. The relative signs of
the oscillating charge density along the metal surfaces are schematically illustrated to the left of the graph.
For $\eta_{\s 0}  =2\pi/5>\pi/4$ (lower panel) the positions of the upper symmetric mode and lower antisymmetric mode with respect to $\omega_{\s 0}$ invert.}
\label{F6}
\end{figure}

In addition to the spectrum, $\omega(m)$,
it is instructive to look at the spatial distribution of electric field of the two plasmon modes.
For a given frequency, $\omega$, the value of $m$ is found by equating it to either $\omega_s(m)$, when $\omega <\omega_{\s 0}/\sqrt{2}$, or to $\omega_a(m)$, when $\omega >\omega_{\s 0}/\sqrt{2}$. The found value of $m(\omega)$ is then substituted into the potential distribution $\Phi_s=\cos(m\xi)\cosh(m\eta)$ or $\Phi_a=\sin(m\xi)\sinh(m\eta)$,
from which the electric field is subsequently calculated.

To clarify the physical meaning of the parameter $m$
we rewrite the potential $\Phi_s$ at large distances from
the coordinate origin in the Cartesian coordinates

\begin{align}
\label{cartesian}
\Phi_s(x,y)\vert_{x,y \gg a}&=C\cos\Biggl[m\Biggl(\ln\frac{2\sqrt{x^2+y^2}}{a}\Biggr)\Biggr]
\nonumber\\
&\times\cosh\Bigl[ m\arctan\frac{y}{x}\Bigr].
\end{align}
The ratio of the amplitudes
of  $\Phi_s$ at the boundary, $y=x\tan \eta_{\s 0}$,
and along the $x$-axis is equal  to $\cosh\left(m\eta_{\s 0}\right)$.

The potential distribution Eq.~(\ref{cartesian}) applies inside the metal.
In the air, this distribution differs from  Eq. (\ref{cartesian}) by the replacement
$\arctan \Bigl( \frac{y}{x}\Bigr)$ by $\pi-\arctan \Bigl(\frac{y}{x}\Bigr)$ and $C$ by
$C\cosh (m\eta_{\s 0})/\cosh \left[m(\pi-\eta_{\s 0})\right]$.

Consider now a symmetric plasmon propagating with a wavevector, $q$, in a metallic film of a constant thickness, $2d$. For this plasmon
the ratio of the potentials at the boundary and at the center is equal
to $\cosh \left(qd\right)$. Identifying $d$ with the length of the arc between the $x$-axis
and the boundary, $\rho\eta_{\s 0}$, where $\rho=\sqrt{x^2+y^2}$, allows one to specify the local value of the wavevector
\begin{equation}
\label{q}
q(\rho)=\frac{m}{\rho}.
\end{equation}
With the $\rho$-dependent wavevector given by Eq. (\ref{q}),
one would expect for the plasmon phase a ``semiclassica'' value
$\int_a^{\rho}d\rho'q(\rho')=m\ln(\rho/a)$, which is indeed the case, as follows from Eq.~(\ref{cartesian}).

For a symmetric mode, the electric field on the $x$-axis is directed
along $x$. The domain $x>a$ on the $x$-axis corresponds to
$\eta=0$, so that $\Phi_s(\xi)=C\cos \left(m\xi\right)$.
In this domain the behavior of electric field with $x=a\cosh \xi$
has the form
\begin{equation}
\label{field1}
 E_{x}(x>a)=Cm~
\frac{\sin \Bigl[ m\ln \Bigl(\frac{x}{a}+\sqrt{\frac{x^2}{a^2}-1}\Bigr)\Bigr]}
{\sqrt{x^2-a^2}}.
\end{equation}
The field is finite at $x=a$, and falls off as  $1/x$ for $x\gg a$.

In the domain $a\cos \eta_{\s 0}<x<a$ we have $\xi=0$ and
$\Phi_s=C\cosh \left(m\eta\right)$. Differentiating with
respect to $x=a\cos \eta$, we obtain
\begin{equation}
\label{field2}
E_{x}(a\cos \eta_{\s 0}< x<a)=Cm~
\frac{\sinh\Bigl( m \arccos\frac{x}{a}\Bigr)}
{\sqrt{a^2-x^2}}.
\end{equation}
Finally, in the air, in the domain $0<x<a\cos \eta_{\s 0}$
the $x$-component of the field is given by
\begin{align}
\label{field2'}
E_{x}(0<x<a\cos \eta_{\s 0})\nonumber \\
=-mC\Biggl(\frac{\cosh(m\eta_{\s 0})}{\cosh\left[m(\pi-\eta_{\s 0})\right]}\Biggr)
\frac{\sinh \left[m(\pi-\arccos\frac{x}{a})\right]}
{\sqrt{a^2-x^2}}.
\end{align}
In Eqs. (\ref{field1}), (\ref{field2}) the frequency and the $\eta_{\s 0}$-dependence of electric field
is incorporated in $m\left(\omega,\eta_{\s 0}\right)$, defined by the condition
$\omega_s (m)=\omega$, where $\omega_s (m)$ is given by Eq. (\ref{single symmetric}).

For an antisymmetric mode, the electric field on the $x$-axis is directed along $y$.
Inside the metal, where $\Phi_a=C\sin (m\xi)\sinh (m\eta)$, the $x$-dependence of $E_{\s y}$ is the same as the $x$-dependence, Eqs. (\ref{field2}), (\ref{field2'}), of $E_{\s x}$
for the symmetric mode. In the air, the $x$-dependence of  $E_{\s y}$ is different from
Eq.~(\ref{field2'}) and has the form

\begin{align}
\label{field3}
E_{y}(0<x<a\cos \eta_{\s 0})\nonumber \\
=-mC\Biggl(\frac{\sinh(m\eta_{\s 0})}{\sinh\left[m(\pi-\eta_{\s 0})\right]}\Biggr)
\frac{\sinh \left[m(\pi-\arccos\frac{x}{a})\right]}
{\sqrt{a^2-x^2}}.
\end{align}
The behavior of $E_{\s x}(x)$ and $E_{\s y}(x)$ is illustrated in Fig.~\ref{fields}a.
It is seen that the oscillating tail emerges only at large distance $\sim 20a$ from the
tip.

\vspace{3mm}

\section{Geometry of two hyperbolas}

\subsection{Splitting of the Plasmon Spectrum}

Consider now the geometry of two hyperbolas, Fig. \ref{asy}.
The boundary of the first hyperbola is defined by the same Eq. (\ref{boundary1}); in addition, metal occupies
the domain $\pi-\eta_{\s 1}<\eta<\pi$.
The $\xi$-dependence of the potential for the symmetric modes remains the same,
$\cos m\xi$. The $\eta$-dependence, which decays away from both
boundaries, is characterized by four constants,
\begin{multline}
\label{double symmetric}
\Phi_s(\eta)\\
=\left\{
\begin{aligned}
&A_1\cosh(m\eta),\quad 0<\eta<\eta_{\s 0}\\
&B_1\cosh(m\eta)+B_2\cosh m(\pi-\eta),~~ \eta_{\s 0}<\eta <\pi-\eta_{\s 1}\\
&A_2\cosh m(\pi-\eta),\quad \pi-\eta_{\s 1}<\eta<\pi,
\end{aligned}
\right.
\end{multline}
These constants, $A_1$, $A_2$, $B_1$ and $B_2$ are related by continuity of $\Phi_s(\eta)$ and $\varepsilon \partial \Phi_s/\partial \eta$ at $\eta=\eta_{\s 0}$ and $\eta =\pi -\eta_{\s 1}$. These continuity conditions read
\begin{align}
\label{continuity1}
A_1\cosh(m\eta_{\s 0})=B_1\cosh(m\eta_{\s 0})+B_2\cosh m(\pi-\eta_{\s 0}), \nonumber\\
\varepsilon(\omega)A_1\sinh(m\eta_{\s 0})=B_1\sinh(m\eta_{\s 0})-B_2\sinh m(\pi-\eta_{\s 0}),
\end{align}

\begin{align}
\label{continuity2}
A_2\cosh(m\eta_{\s 1})=B_1\cosh m(\pi-\eta_{\s 1})+B_2\cosh (m\eta_{\s 1}), \nonumber\\
-\varepsilon(\omega)A_2\sinh(m\eta_{\s 1})=B_1\sinh m(\pi-\eta_{\s 1})-B_2
\sinh (m \eta_{\s 1}).
\end{align}
These relations, together with the explicit form of $\varepsilon(\omega)$, given in Eq. (\ref{epsilon}), lead to the following characteristic equation

\begin{widetext}
\begin{equation}
\label{quadratic equation}
\frac{\sinh 2m\eta_{\s 0}\sinh 2m\eta_{\s 1}}{4}
=\Biggl[\Bigl(\frac{\omega^2}{\omega_{\s 0}^2}-\frac{1}{2}\Bigl)\sinh m\pi+\frac{1}{2}\sinh m(\pi-2\eta_{\s 0})\Biggr]
\Biggl[\Bigl(\frac{\omega^2}{\omega_{\s 0}^2}-\frac{1}{2}\Bigl)\sinh m\pi+\frac{1}{2}\sinh m(\pi-2\eta_{\s 1})\Biggr].
\end{equation}
\end{widetext}
The two brackets in the right-hand side describe the  plasmon dispersion
Eq. (\ref{single symmetric}) for a symmetric mode, while the left-hand
side describes the coupling of the two plasmon branches.
The two plasmons decouple when $\eta_{\s 1}$ is small.
Then the dispersion of the upper symmetric branch is simply $\omega_s(\eta_{\s 0},m)$,  Eq. (\ref{single symmetric}).
In order to find  the dispersion of the lower symmetric
branch, it is sufficient to set $\omega=0$ in the first bracket.
This yields
\begin{equation}
\label{small}
\omega_s^{-}(m)\Big|_{\eta_{\s 1}\ll 1}=\omega_{\s 0}
\Bigl[m\eta_{\s 1}\tanh m(\pi-\eta_{\s 0})\Bigr]^{1/2}.
\end{equation}

The general expression for the dispersion of the two coupled symmetric branches reads

\begin{widetext}
\begin{eqnarray}
\label{solution}
\omega_s^{\pm}(m)=
\frac{\omega_{\s 0}}{\bigl[2\sinh (m\pi)\bigr]^{1/2}}\Biggl\{\Bigl[\sinh (m\eta_{\s 0})\cosh m(\pi-\eta_{\s 0})+\sinh (m\eta_{\s 1})
\cosh m(\pi-\eta_{\s 1}) \Bigr]\nonumber\\
\pm \Bigl[\sinh^2 m\left(\eta_{\s 0}-\eta_{\s 1}\right)\cosh^2 m\left(\pi-\eta_{\s 0}-\eta_{\s 1}\right)+
\sinh\left(2m\eta_{\s 0}\right)\sinh\left(2m\eta_{\s 1}\right)\Bigr]^{1/2}\Biggr\}^{1/2}.
\end{eqnarray}
\end{widetext}
It is easy to see from Eq. (\ref{solution}) that for large $m  \gg 1$ both frequencies approach the surface plasmon frequency $\omega_{\s 0}/\sqrt{2}$, as in the case of a single hyperbola. The reason is that the short-wavelength  plasmon is ``local,'' a metal surface is locally flat, and the presence of the second surface is of no consequence to the spectrum in this limit.
The behavior of $\omega_s^{\pm}(m)$ in the limit of long wavelengths is remarkable
\begin{eqnarray}
\label{long wavelength}
&& \omega_s^{-}(m)\Big|_{m\rightarrow 0}\approx m\omega_{\s 0}
\Biggl[\frac{\eta_{\s 0}\eta_{\s 1}\left(\pi-\eta_{\s 0}-\eta_{\s 1}\right)}{\eta_{\s 0}+\eta_{\s 1}}\Biggr]^{1/2}, \nonumber \\
&& \omega_s^{+}(0)=\omega_{\s 0}\Bigl(\frac{\eta_{\s 0}+\eta_{\s 1}}{\pi}   \Bigr)^{1/2}.
\end{eqnarray}
The fact that $\omega_s^{+}(0)$ is determined by the ``net" angle
$(\eta_{\s 0}+\eta_{\s 1})$  is consistent with the result
Eq. (\ref{longwave}) for a single hyperbola. As this net angle approaches $\pi$, the portion of air in this
limit becomes small, and  $\omega_s^{+}(0)$, approaches the
bulk plasmon frequency.  The acoustic behavior of $\omega_s^{-}(m)$ is related to the fact that, unlike for a single hyperbola,  in the geometry of two hyperbolas a low-frequency plasmon is not reflected from the tip, but goes ``through'' the gap into the second hyperbola.

For two identical hyperbolas, $\eta_{\s 0}=\eta_{\s 1}$, Eq. (\ref{solution})
simplifies to
\begin{multline}
\label{double symmetric1}
\omega_s^{\pm}(m)\\
=\omega_{\s 0}\Biggl\{ \frac{\sinh(m\eta_{\s 0})
\Bigl(\cosh \left[m(\pi-\eta_{\s 0})\right]\pm \cosh (m\eta_{\s 0})\Bigr)}{\sinh(m\pi)}\Biggr\}^{1/2}.
\end{multline}
Similar derivation for the antisymmetric plasmons yields

\begin{multline}
\label{double asymmetric}
\omega_a^{\pm}(m)\\
=\omega_{\s 0}\Biggl\{ \frac{\cosh(m\eta_{\s 0})
\Bigl(\sinh \left[m(\pi-\eta_{\s 0})\right]\pm \sinh (m\eta_{\s 0})\Bigr)}{\sinh(m\pi)}\Biggr\}^{1/2}.
\end{multline}
From Eqs. (\ref{double symmetric}), (\ref{double asymmetric})
one can trace the evolution of the plasmon spectrum with increasing $\eta_{\s 0}$. For $\eta_{\s 0} \ll 1$ the frequencies of both
symmetric plasmons are low:

\begin{equation}
\label{asymptote1}
\omega_s^{-}(m)\Big|_{\eta_{\s 0}\ll 1}\approx \omega_{\s 0}
\Biggl[m\eta_{\s 0}\tanh\left(\frac{m\pi}{2}\right)\Biggr]^{1/2},
\end{equation}

\begin{equation}
\label{asymptote2}
\omega_s^{+}(m)\Big|_{\eta_{\s 0}\ll 1}\approx \omega_{\s 0}
\Biggl[\frac{m\eta_{\s 0}}{\tanh\left(\frac{m\pi}{2}\right)}\Biggr]^{1/2},
\end{equation}
while the frequencies of antisymmetric plasmons are close to $\omega_{\s 0}$. As $\eta_{\s 0}$ increases and achieves the value $\pi/4$, the branches $\omega_s^{+}(m)$ and $\omega_a^{-}(m)$ collapse into a single frequency $\omega_{\s 0}/\sqrt{2}$ and become flat.
As $\eta_{\s 0}$ increases further above $\pi/4$, the branches invert:  $\omega_s^{+}(m)$ is pushed above $\omega_{\s 0}/\sqrt{2}$  and $\omega_a^{-}(m)$ drops below it. This evolution is illustrated in
Fig. \ref{F6}.

For the geometry of two hyperbolas there is a duality relation,
\begin{equation}
\label{duality2}
\omega_s^{+}\left(\frac{\pi}{2}-\eta_{\s 0},m\right)=
\omega_a^{-}\left(\eta_{\s 0},m\right),
\end{equation}
similar to Eq. (\ref{duality}) for a single hyperbola.


\vspace{3mm}

\subsection{Comparison of the field distributions for two geometries}

 Hybridization of plasmon fields of individual hyperbolas
in the geometry of two hyperbolas is illustrated
in Fig. \ref{fields}, Fig.~\ref{F4}, and Fig.~\ref{F5}.
The curves in Fig. \ref{fields} suggest that
 hybridization {\em along the} $x$-{\em axis} is rather weak
and becomes progressively weaker while the wavenumber
increases as the frequency approaches $\omega_{\s 0}/\sqrt{2}$.
This behavior is natural, since, the closer the frequency is to that of the flat  surface
plasmon, the more localized is the plasmon field near the metal-air
interface. The density plot of the modes of individual hyperbolas is shown in Fig.~\ref{F4}.
Compared with the latter, the double-hyperbola plot of Fig.~\ref{F5} demonstrates that hybridization
of individual modes has a ``ring''-like character
for symmetric modes and the ``needle''-like character for antisymmetric modes.
Such different nature of hybridization can be interpreted with the help of the patterns of oscillating surface charges in Fig. \ref{F6}.
For the bottom symmetric mode with  low frequency, the positive and negative charges are separated by
air, whereas the dielectric function of the metal is large. This expels the electric field lines from the inside of the metal and forces them to  go through the air. The field is strong along the $y$-axis, where it is parallel to the $x$-axis.

For the antisymmetric mode, the upper and lower  sides of the metal have opposite charges,
while the  frequency is close to $\omega_{\s 0}$, so that the dielectric constant of the
metal is small. The force lines of electric field are localized inside the metal. Thus, the two metal edges can be viewed as the plates of a  parallel-plate capacitor.
Correspondingly, the concentration of electric field
near the $x$ axis is analogous to the
fringe field outside a parallel-plate capacitor.

In Sect.~II we have established that the field of the plasmon modes at large distance,
$\rho \gg a$, from the origin behaves as $1/\rho$. Hence, the field intensity
behaves as $1/\rho^2$, i.e. it strongly diverges down to the distances $\sim a$.
This, however, does not translate into a strong enhancement of the {\em net} energy,
$\int d{\bf r}\,{\bf E}^2({\bf r})$, which grows only logarithmically,
$\propto \ln(\lambda/a)$; the upper cut-off  is provided by the wavelength of light, $\lambda =2\pi c/\omega$,
with the same frequency $\omega$.

\subsection{Two Co-directed Hyperbolas}

\begin{figure}[t]
\centerline{\includegraphics[width=90mm,angle=0,clip]{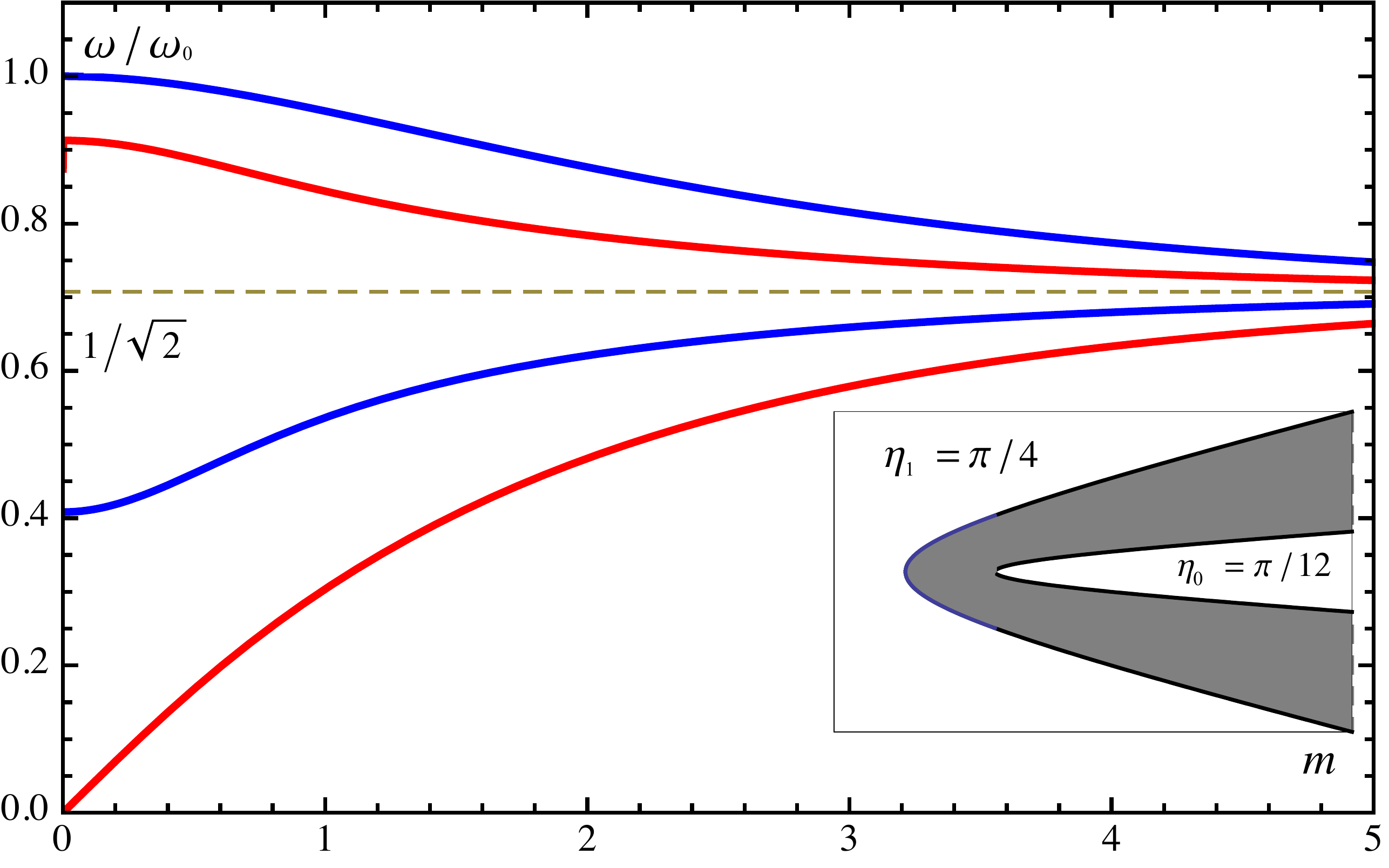}}
\caption{(Color online) Illustration of the plasmon spectrum in the geometry of two co-directed hyperbolas. Blue and red curves correspond to symmetric and antisymmetric plasmon modes, respectively. The spectrum is plotted from Eq. (\ref{quadratic equation1}) for $\eta_c=\delta\eta=\pi/6$. }
\label{F7}
\end{figure}

For completeness, in this subsection we will analyze the plasmon spectrum in the geometry of two co-directed hyperbolas. Assume that the metal occupies the region $\eta_{\s 0}<\eta <\eta_{1}$, while the regions $0<\eta<\eta_{\s 0}$ and $\eta_{\s 1}<\eta <\pi$ are occupied by air, see the inset in Fig.~\ref{F7}. A straightforward generalization of Eq. (\ref{quadratic equation}) gives

\begin{widetext}
\begin{equation}
\label{quadratic equation1}
\frac{\sinh 2m\eta_{\s 0}\sinh 2m\left(\pi-\eta_{\s 1}\right)}{4}
=\Biggl[\Bigl(\frac{\omega^2}{\omega_{\s 0}^2}-\frac{1}{2}\Bigl)\sinh m\pi-\frac{1}{2}\sinh m(\pi-2\eta_{\s 0})\Biggr]
\Biggl[\Bigl(\frac{\omega^2}{\omega_{\s 0}^2}-\frac{1}{2}\Bigl)\sinh m\pi+\frac{1}{2}\sinh m(\pi-2\eta_{\s 1})\Biggr].
\end{equation}
\end{widetext}
To analyze the plasmon dispersion, we introduce the average opening angle and the ''thickness'' of the tip,
\begin{equation}
\eta_c=\frac{\eta_{\s 0}+\eta_{\s 1}}{2},~~~~~~~~\delta\eta=\eta_{\s 1}-\eta_{\s 0}.
\end{equation}


An example of the plasmon spectrum with the two co-directed hyperbolas is shown in Fig.~\ref{F7}. In this inverted geometry, symmetric and antisymmetric modes interchange, compared with Fig.~\ref{F6}(b). The symmetric branches in the long-wavelength limit are
\begin{equation}
\omega_s^{+}(0)=\omega_{\s 0}\Bigl(\frac{\delta \eta}{\pi}\Bigr)^{1/2},\hspace{0.4cm} \omega_s^{-}(0)=\omega_{\s 0}.
\end{equation}

For the upper symmetric mode, $\omega_s^{+}(0)$, the signs of the oscillating charges on the opposite sides of each
``sleeve''
are opposite and the  electric field lines are confined inside the
metal. For this,  the dielectric constant of the metal must vanish. For the lower symmetric mode, $\omega_s^{-}(0)$, the signs of charges on the opposite sides of each
sleeve
are the same. The field lines mostly stay in the air and do not penetrate into the metal. This, on the other hand, implies that $\varepsilon(\omega)$ is large, and, correspondingly, the frequency of the mode is small, vanishing in the limit of a very thin ``coating,'' $\delta \eta \to 0$. 

The acoustic mode,  generic for multi-connected geometries, is now found in the antisymmetric part of the spectrum, $\omega_a^{+}(m) \to 0$ as $m\to 0$. The second mode,
\begin{equation}
\omega_a^{-}(0)=\omega_{\s 0}\Bigl(1-\frac{\delta \eta}{\pi}\Bigr)^{1/2}.
\end{equation}
lies above $\omega_{\s 0}/\sqrt{2}$ and its frequency increases with decreasing $\delta \eta$. 

\vspace{5mm}

\begin{figure}[t]
\centerline{\includegraphics[width=90mm,angle=0,clip]{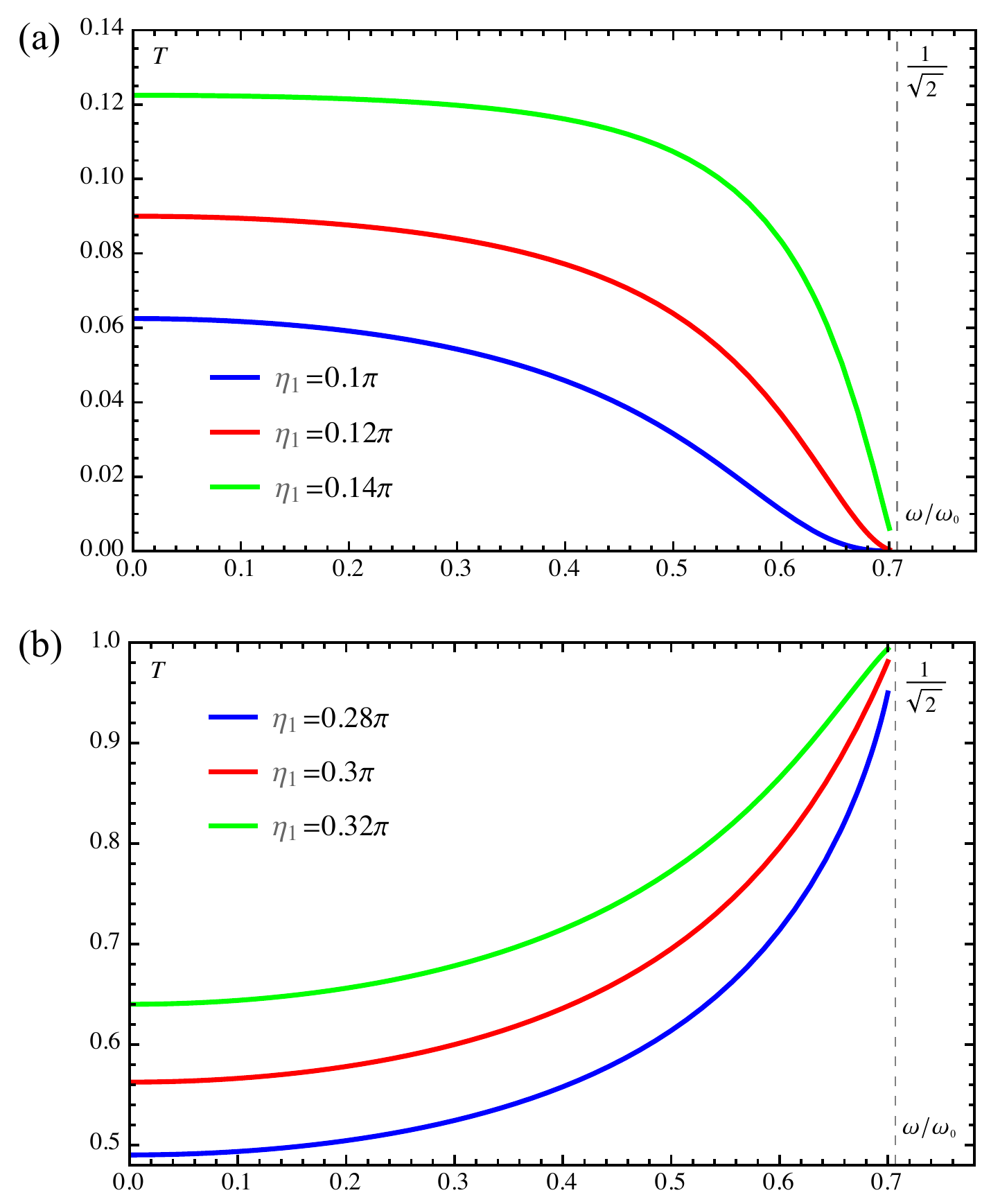}}
\caption{(Color online) Geometric factor for the energy transfer is plotted as a function of frequency from Eq. (\ref{geometrical1}) for the values of the opening angle $2\eta_{\s 1}=0.2\pi,~0.24\pi,~0.28\pi$ (a),
and $2\eta_{\s 1}=0.56\pi,~0.6\pi,~0.64\pi$ (b). The  value of the opening angle, $2\eta_{\s 0}$, is chosen $2\eta_{\s 0}=4\pi/5$. }
\label{F10}
\end{figure}

\section{Interaction of two emitters at the metal-air surface}

Numerous studies have demonstrated that, due to proximity to the metal, the lifetime of the emitter can be dominated by the excitation of the plasmon modes
\cite{SPPEnhanced,lifetime1,lifetime2,lifetime3,GovorovEnhanced,Cooperative,Zhenya,Lukin1,Lukin2}.
For a metallic nanoparticle  plasmons significantly shorten the lifetime when the emitter frequency
is close to the frequency of the plasmon dipole ($l=1$) oscillations. A less trivial finding\cite{Lukin1,Lukin2} is that the coupling to plasmons can dominate the lifetime when the emitter is located in the proximity to a nanowire with a subwavelength radius in which the plasmon spectrum is continuous.
When the emitter is positioned close to the tip, the plasmon-induced shortening of the
lifetime is even more pronounced\cite{Lukin2}. This is caused by the field enhancement near the tip.
In this regard, to calculate the emitter lifetime in a multi-connected plasmonic structure, we can use the same general scheme as developed in the previous studies.

Consider for concreteness, a dipole emitter with frequency $\omega$,
located at the metal-air interface, and polarized parallel
to the interface, as in Fig. \ref{asy}.
The structure of the plasmon spectrum
in this calculation is captured by  Green's function for electric field,
\begin{equation}
\label{Greens function}
G(\xi,\eta; \xi',\eta';\omega)=\sum_m \frac{E_{\s \xi}(\xi,\eta;m) E_{\s \xi}(\xi',\eta';m)}{\omega-\omega(m)},
\end{equation}
where $E_{\s \xi}(\xi,\eta;m)$ is the normalized tangential component of the electric
field of the plasmonic mode and the summation is taken over all four modes;
$\omega(m)$ is the plasmon spectrum found above,
The decay rate is proportional to the imaginary part
of the diagonal value $G (\xi,\eta_i; \xi,\eta_i;\omega)$ taken at the position of emitter.
As the summation  in Eq. (\ref{Greens function})   is performed
over  acoustic as well as optical modes, a spike-like feature occurs in the $\omega$-dependence of
the decay rate. The origin of this feature is the divergence of the
density of the plasmon modes,
\begin{equation}
\label{density}
g^{+}(\omega)=\frac{1}{\pi}\int_0^{\frac{\omega_{\s 0}}{\sqrt{2}}} dm~\delta\bigl[\omega-\omega^{+}(m)\bigr],
\end{equation}
near the threshold frequency $\omega_s^{+}(0)$.

Consider for simplicity a symmetric geometry, $\eta_{\s 0}=\eta_{\s 1}<\pi/4$.
Expanding Eq. (\ref{double symmetric}) at small $m$, we get
\begin{equation}
\label{expansion}
\frac{\omega_s^{+}(m)}{\omega_s^{+}(0)}\Big|_{m\rightarrow 0}=
1+\frac{m^2}{3}\Bigl(\frac{\pi}{4}-\eta_{\s 0}\Bigr)\Bigl(\frac{\pi}{2}-\eta_{\s 0}\Bigr),
\end{equation}
where $\omega_s^{+}(0)=\left(2\eta_{\s 0}/\pi\right)^{1/2}$. Substituting Eq. (\ref{expansion}) into Eq. (\ref{density}), one finds
\begin{equation}
\label{density}
g^{+}(\omega)=\frac{1}{\pi}\Biggl[\frac{3}{\left(\frac{\pi}{4}-\eta_{\s 0}\right)\left(\frac{\pi}{2}-\eta_{\s 0}\right)\omega_s^{+}(0)\left(\omega-\omega_s^{+}(0)\right)}\Biggr]^{1/2}.
\end{equation}
The inverse square-root singularity in the density of states translates into a
minimum in the emitter lifetime. Similar minimum is expected near
$\omega =\omega_a^{+}(0)$ and near the bulk plasmon frequency, $\omega_{\s 0}$.
Near $\omega =\omega_{\s 0}/\sqrt{2}$ the minimum should be even stronger, since the
divergence of the plasmon modes diverges as $\vert\omega -{\omega_{\s 0}}/{\sqrt{2}}\vert^{-1}$.

The scheme of calculation of the plasmon-mediated energy transfer
between the emitters  near the metal-air interface is also well established,
see e.g. Ref. \onlinecite{GovorovEnhanced} for two emitters near a nanoparticle
and Ref. \onlinecite{Velizhanin} for donor and acceptor above the graphene.
Here we emphasize the specifics of this transfer in the geometry of two
hyperbolas. In particular, we study how the transfer rate between the red emitter
near the tip in Fig. \ref{asy} and the blue emitter located on the left hyperbola
differs from the transfer rate between the red emitter and the blue emitter
located on the right hyperbola, when both blue emitters have the same coordinate, $\xi$.

Quantitatively, the relation of the two transfer rates is described by the ratio
of the tangential components of  electric field at the two interfaces. From Eq. (\ref{double symmetric}) we find
\begin{eqnarray}
\label{geometrical}
&&T=\Biggl(\frac{E_{\xi}(\eta_{\s 0},\omega)}{E_{\xi}(\eta_{\s 1},\omega)}\Biggr)^2\nonumber \\
&&=
\Biggl[\frac{B_1\cosh(m\eta_{\s 0})+B_2\cosh m(\pi-\eta_{\s 0})}{B_1\cosh m(\pi-\eta_{\s 1})+B_2\cosh (m\eta_{\s 1})}\Biggr]^{2}.
\end{eqnarray}
With the help of the continuity conditions  Eqs. (\ref{continuity1}), (\ref{continuity2}) this ratio can be cast in the form

\begin{equation}
\label{geometrical1}
T=\Biggl[\varepsilon(\omega)\tanh (m\eta_{\s 1})\sinh m(\eta_{\s 0}-\eta_{\s 1})
+\cosh m(\eta_{\s 1}-\eta_{\s 0})\Biggr]^{-2},
\end{equation}
where $m(\omega)$ is determined by the dispersion equation (\ref{solution}).
In Fig.~\ref{F10} we plot the factor $T$ as function of frequency for
a fixed opening angle of the right hyperbola, $2\eta_{\s 0}=4\pi/5$, and
several opening angles of the left hyperbola.
Overall, we see that at $\omega =0$ the geometric factor $T$ rapidly increases with
$\eta_{\s 1}$. This can be understood from the electrostatics of ideal metals. Indeed, at $\omega=0$
the field does not penetrate into the metal at all. Consequently, the sharper the left hyperbola is,
the stronger  the field near the left tip becomes, resulting in smaller values of $T(0)$.  This, in turn, means that the energy transfer from the red emitter in Fig. \ref{asy} to the left blue emitter happens faster than the transfer to the right blue emitter.

We also see that the behavior of $T(\omega)$ is different for small, Fig.~\ref{F10}a), and large, Fig.~\ref{F10}b), values of $\eta_{\s 1}$. When geometrical factor $T(0)$ is small, $T(\omega)$ falls off with frequency, suggesting that the left and right hyperbolas become effectively ``disconnected''. To the contrary, for larger starting values $T(0)$, the finite-frequency $T(\omega)$ grows with frequency, suggesting that the energy transfer becomes more symmetric.  The crossover from decay to growth takes place at $\eta_{\s 1}\approx 0.19\pi$. This value can be related to the peculiar behavior of the velocity of the low-frequency plasmon Eq. (\ref{long wavelength}). Namely, this velocity
has a minimum as function of $\eta_{\s 1}$. By setting $\partial \omega_s^{-}/\partial \eta_{\s 1}=0$, we  find the position of this minimum at,
\begin{equation}
\label{tilde}
{\tilde \eta}_{\s 1}=\left(\pi\eta_{\s 0}\right)^{1/2}-\eta_{\s 0}.
\end{equation}
For $\eta_{\s 0}=2\pi/5$, Eq. (\ref{tilde}) yields
${\tilde \eta}_{\s 1} \approx 0.23\pi$, close to the crossover value of
$\eta_{\s 1}$, which corresponds to $\partial T(\omega)/\partial \omega \vert_{\omega\rightarrow 0} =0$.

\section{Concluding remarks}

(i) The prime qualitative finding of the present manuscript is that, for a singly-connected geometry,
the opening angle of the hyperbola, $2\eta_{\s 0}$, defines the minimum frequency, given by Eq. (\ref{longwave}), below which the plasmon bound to the tip of the hyperbola does not exist.
 By contrast,
in a multi-connected geometry a symmetric plasmon mode, concentrated near the region of the closest contact of the surfaces, exist at arbitrary low frequency.

(ii) Originally, the enhancement of the electric field of
a plasmon-polariton as it approaches a wedge, has been demonstrated analytically
in Ref. \onlinecite{original1}. Calculation in Ref. \onlinecite{original1},
was carried out in polar coordinates, i.e. neglecting  the rounding
of the wedge near the tip. This curving has been emulated by introducing a cutoff
length $\sim a$. At distances much greater than $a$ the form of the
plasmon field in Ref. \onlinecite{original1} is consistent with  Eq. (\ref{cartesian}) with one important difference that the
plasmon field Eq. (\ref{cartesian}) is not traveling,
as in Ref. \onlinecite{original1}, but is a standing wave instead.
This structure is enforced by the boundary condition at the tip.

The fact that the plasmon mode near the tip is a standing wave can
be viewed from the perspective of focusing  of light on
subwavelength scale,
when the plasmon-polariton with TM polarization (magnetic field
along the $z$-axis) is excited at large (compared to the wavelength, $\lambda$) distance, $\rho$, from the tip\cite{Stockman}. In the geometry of a single hyperbola
this polariton will be fully reflected. For small enough opening angles,
$2\eta_{\s 0}$, the transformation of the polariton into small-$\rho$ standing plasmon
mode can be traced
 analytically within the semiclassical (WKB)
description\cite{Stockman,AfterStockman0,AfterStockman1,JAP,Lukin1,Lukin2}. Within
that description, the metal strip at distances $\rho \gg a$ from the origin is
replaced with a planar film with a thickness $d_{\s \rho}$, equal to the arc distance,
 $2\rho\eta_{\s 0}=d_{\s \rho}$, between the two metal surfaces.
 Then the semiclassical expression
for the incident and reflected fields has the form
${q_{\s \rho}^{-1/2}}\exp\left[\pm i \int d\rho~q_{\s \rho}\right]$, where
$q_{\s \rho}(\omega)$ is the dispersion of the symmetric waveguide mode propagating
along the film. This dispersion law satisfies the equation

\begin{equation}
\label{planar}
\varepsilon\tanh\Biggl\{\Biggl(q_{\s \rho}^2-\varepsilon\frac{\omega^2}{c^2}\Biggr)^{1/2}\frac{d_{\s \rho}}{2}\Biggr\}=-\Biggl(\frac{q_{\s \rho}^2-\varepsilon\frac{\omega^2}{c^2}}{q_{\s \rho}^2-\frac{\omega^2}{c^2}}
       \Biggr)^{1/2}.
\end{equation}
The relevant question is, how close to the threshold frequency
$\omega_{\s 0}\Bigl(\eta_{\s 0}/\pi\Bigr)^{1/2}$  is the semiclassical
description applicable, i.e. when does the condition $dq_{\s \rho}/d\rho \ll q_{\s \rho}^2$
apply? At small frequencies we can replace
$\varepsilon(\omega)$ by  $-\omega_{\s 0}^2/\omega^2$ and simplify Eq. (\ref{planar})
using the identification $q_{\s \rho}=m/\rho$, Eq. (\ref{q}). Then Eq. (\ref{planar})
reduces to the following equation for $m$
\begin{equation}
\label{planar1}
\tanh \Biggl[\eta_{\s 0}m\Bigl(1+\frac{\omega_{\s 0}^2\rho^2}{c^2m^2}\Bigr)^{1/2}     \Biggr]=\frac{\omega^2}{\omega_{\s 0}^2}\Bigl(1+\frac{\omega_{\s 0}^2\rho^2}{c^2m^2}\Bigr)^{1/2},
\end{equation}
and the semiclassical condition is satisfied provided that $m\gg 1$.
As the distance $\rho$ increases, the solution of Eq. (\ref{planar1})
grows starting from $m=\frac{\omega^2}{\omega_{\s 0}^2\eta_{\s 0}}$, which is
consistent with
Eq. (\ref{single symmetric}).
Thus, for semiclassical description to apply at all $\rho$ bigger than $a$, it
is necessary that this minimum $m$ exceeds $1$, i.e.
$\omega \gg \omega_{\s 0}\eta_{\s 0}^{1/2}$. The latter condition suggests that
the frequency of the incident light, being much smaller than $\omega_{\s 0}$,
should still not be very close to the threshold frequency.

Another approach to the  ``delivery" of the light energy to the tip
was proposed in Ref. \onlinecite{AfterStockman2}. Namely, one can
coat a conical tip of a glass fiber by a silver layer, the geometry
similar to the one shown in the inset in Fig.~\ref{F7}. The wavelength of a plasmon in a silver
layer increases away from the tip, and at a certain distance matches the
wavelength of the waveguided light propagating towards the tip.
At this point the light transforms into the plasmon and heads towards
the tip. Experimental realization of a coupling based
on this idea was reported in a number of papers,
see Ref. \onlinecite{NumericalHyperbolas} and references therein.

(ii) Naturally, the structure of the plasmon field at small distances from
the tip is not captured in polar coordinates.
Meanwhile, this structure
is quite nontrivial. To illustrate this, consider the ratio $E_{\s x}(0)/E_{{\s x},tip}$ of the fields at the origin and at the tip for a symmetric plasmon.  This ratio can be
found from Eq. (\ref{field2}). The origin, $x=0$, corresponds to $\eta=\pi/2$, while the tip corresponds to $x=a\cos\eta_{\s 0}$. Then Eq. (\ref{field2}) yields
\begin{equation}
\label{ratio}
\frac{E_{\s x}(0)}{E_{{\s x},tip}}=\frac{\sin\eta_{\s 0}\sinh\left[\frac{\pi}{2}m(\omega)\right]}
{\sinh\left[(\pi-\eta_{\s 0})m(\omega)\right]}.
\end{equation}
As the frequency grows from the threshold value
$\omega_s(0)=\omega_{\s 0}\Bigl(\eta_{\s 0}/\pi\Bigr)^{1/2}$ to the surface plasmon frequency $\omega_{\s 0}/\sqrt{2}$, the ``wavenumber,'' $m$,
changes from $m=0$ to $m=\infty$. Then the ratio Eq. (\ref{ratio})
falls off from $\pi\sin \eta{\s 0}/[2\left(\pi-\eta_{\s 0}\right)]$
monotonically.
At frequencies close to $\omega_{\s 0}/\sqrt{2}$ the ratio Eq. (\ref{ratio})
approaches zero, since the plasmon field is strongly localized near
the metal surface.
Such a strong change of the field is revealed by the exact
solution in elliptic coordinates demonstrated in the present paper.

(iii) In this paper we assumed that the plasmon field does not depend on $z$.
For a general $z$-dependence, $\Phi \propto \exp(i\kappa z)$,  the Laplace equation
takes the form

\begin{equation}
\label{LaplaceWithZ}
\frac{\partial^2\Phi}{\partial \xi^2}+\frac{\partial^2\Phi}{\partial \eta^2}
+\kappa^2a^2(\cosh^2\xi -\cos^2\eta)\Phi=0.
\end{equation}
Separation of variables, leads to the following equation for the potential's $\eta$-dependence,
\begin{equation}
\frac{\partial^2\Phi}{\partial \eta^2}-(m^2+\kappa^2a^2\cos^2\eta)\Phi=0.
\end{equation}

Our results apply for the wavenumbers that exceed some minimum value,  $m\gg m_{\s min}$. This minimum value can be estimated from the observation that $\kappa$ can not be less than $H^{-1}$, where $H$ is the ``height" of the hyperbola, see Fig.~\ref{F1}. This determines,
$m_{\s min} \sim \kappa a \sim a/H \ll 1$.

(iiii) A finite scattering rate of conduction electrons, $\gamma$,  sets the maximum distance from the origin, $\rho_{\s max}$, where our results apply. The solutions obtained and discussed above decay beyond  $\rho_{\s max}$.
This distance can be estimated by noting that a finite $\gamma$ is taken into account by replacing $\omega^2 \to \omega(\omega + i/\gamma)$ in the dielectric function, Eq.~(\ref{epsilon}). As a result, it leads to a finite imaginary part of $m$: $\text{Im}\,m \approx \gamma \left(\partial \omega/\partial m \right)^{-1}$, where $\partial \omega/\partial m$ is the slope of the plasmon dispersion. The plasmons attenuate at distances where $ \xi\, \text{Im}\, m\sim 1$. Since at large distances $\xi$ depends logarithmically on $\rho$, see Eq. (\ref {cartesian}), we conclude that the maximum distance is
$\rho_{\s max} \sim a \exp{\left[1/\text {Im}~m\right]}$.


\acknowledgements

L.S. and E.M. were supported by the Department  of  Energy,
Office  of  Basic  Energy  Sciences, Grant No. DE-FG02-06ER46313.
M.R. was supported by the NSF MRSEC program at
the University of Utah under Grant No. DMR 1121252.


\end{document}